\documentclass[aps,prr,floatfix,twocolumn,superscriptaddress,showpacs,preprintnumbers,longbibliography,nofootinbib]{revtex4-2}

\usepackage{soul}
\usepackage{setspace}
\usepackage{dcolumn}    
\usepackage{bm} 
\usepackage{graphicx}
\usepackage{amsmath}  
\usepackage{amsthm}   
\usepackage{enumitem}
\usepackage{latexsym}
\usepackage{amsfonts}   
\usepackage{amssymb}
\usepackage{array}      
\usepackage{epsfig}
\usepackage{braket} 
\usepackage{bbold}
\usepackage{color}

\usepackage[colorlinks=true,linkcolor=blue,urlcolor=blue,citecolor=blue,pdfusetitle]{hyperref}
\usepackage{xurl}

\usepackage{ulem}

\usepackage{tikz}
\newcommand{\ra}{\rangle}
\newcommand{\la}{\langle}

\newcommand{\QED}{\hfill\ensuremath{\square}}
\newcommand{\eq}{Eq.~}

\newcommand{\fig}{Fig.~}

\newcommand{\tys}{\hspace{0.5 pt}}

\DeclareRobustCommand\openzero{\leavevmode\hbox{0\kern-.55em0}}

\newcommand{\Ao}[1]{\hat{a}_{#1}}
\newcommand{\Aod}[1]{\hat{a}^\dag_{#1}}

\newtheorem{property}{Property}
\newtheorem{definition}{Definition}
\newtheorem{lemma}{Lemma}

\newtheorem{example}{Example}[section]
\newcommand{\gs}{\tikz[baseline , yshift=2pt]{\node[draw=black, fill=green, minimum size=8pt, anchor=base] {}; }}
\newcommand{\ws}{\tikz[baseline, yshift=2pt]{\node[draw=black, fill=white, minimum size=8pt, anchor=base] {}; }}

\newcommand{\gws}{%
  \tikz[baseline , yshift=2pt]{%
    \node[minimum size=8pt, inner sep=0pt, anchor=base] (box) {};
    \begin{scope}
      \clip (box.south west) rectangle (box.north east);
      \fill[green] (box.south west) -- (box.north east) -- (box.north west) -- cycle;
    \end{scope}
    \draw[black] (box.south west) rectangle (box.north east);
  }%
}

\newcommand{\is}{\tikz[baseline, yshift=2pt]{\node[draw=none, fill=none, minimum size=8pt, anchor=base] {}; }}

\date{\today}

\begin{document}

\author{Dario Cilluffo}
\email{dario.cillufo@uni-ulm.de}
\affiliation{Institute of Theoretical Physics, Ulm University, Albert-Einstein-Allee 11, 89081 Ulm, Germany}
\affiliation{Center for Integrated Quantum Science and Technology (IQST), 89081 Ulm, Germany}
\author{Matthias Kost}
\affiliation{Institute of Theoretical Physics, Ulm University, Albert-Einstein-Allee 11, 89081 Ulm, Germany}
\affiliation{Center for Integrated Quantum Science and Technology (IQST), 89081 Ulm, Germany}
\author{Nicola Lorenzoni}
\affiliation{Institute of Theoretical Physics, Ulm University, Albert-Einstein-Allee 11, 89081 Ulm, Germany}
\affiliation{Center for Integrated Quantum Science and Technology (IQST), 89081 Ulm, Germany}
\author{Martin B. Plenio}
\email{martin.plenio@uni-ulm.de}
\affiliation{Institute of Theoretical Physics, Ulm University, Albert-Einstein-Allee 11, 89081 Ulm, Germany}
\affiliation{Center for Integrated Quantum Science and Technology (IQST), 89081 Ulm, Germany}

\begin{abstract}
Tensor network formalisms have emerged as powerful tools for simulating quantum state evolution. While widely applied in the study of optical quantum circuits, such as Boson Sampling, existing tensor network approaches fail to address the complexity mismatch between tensor contractions and the calculation of photon-counting probability amplitudes. Here, we present an alternative tensor network framework that exploits the input-output relations of quantum optical circuits encoded in the unitary interferometer matrix. Our approach bridges the complexity gap by enabling the computation of the permanent —central to Boson Sampling— with the same computational complexity as the best known classical algorithm based on a graphical representation of the operator-basis MPS that we introduce. Furthermore, we exploit the flexibility of tensor networks to extend our formalism to incorporate partial distinguishability and photon loss, two key imperfections in practical interferometry experiments. This work offers a significant step forward in the simulation of large-scale quantum optical systems and the understanding of their computational complexity.
\end{abstract}

\title{Heisenberg picture tensor network formalism for optical circuits}
\maketitle

The tensor network formalism for the time-evolution of quantum states \cite{Orus_2019,montangero2018introduction,Schollwoeck_2011} offers numerical speedup primarily due to two key factors: the efficient representation of quantum states and operators with bounded correlations \cite{audenaert2002,eisert2010} through a modular structure, and the ability to approximate general quantum states with controlled, quantifiable error. These features lead to significant savings in computational resources \cite{PhysRevLett.91.147902,Or_s_2014}.
They are also the main reason tensor networks have found widespread application in the description of optical quantum circuits, particularly in the context of Boson Sampling \cite{aaronson2011computational,GarciaPatron,oh2024classical}.
In such applications, the evolution of an optical input state through an interferometer is modeled as the propagation of a bosonic 
MPS through a network of Matrix Product Operators (MPO) representing the phase shifters and beam splitters that define the interferometer.
However, when applying these methods to Fock states propagating through such an interferometer -- such as in reproducing ideal Boson Sampling experiments -- the computational scaling of tensor network algorithms and the intrinsic complexity of photon-counting tasks scale very differently. In analogy with the fermionic case \cite{PhysRevLett.102.057202,clark2010exact}, it has been shown \cite{Cilluffo2024boson} that for any defectless passive optical network a product of $n$ bosonic ladder operators evolves as an MPO with a maximum bond dimension of $D = 2^n$, independent of the number of input modes $M$. However, the compression and manipulation of MPOs involve singular value decompositions (SVDs) with a computational cost scaling as $\mathcal{O}(D^3)$ \cite{Oh_2021}. In contrast, photon detection amplitudes can be computed by evaluating the permanent of a special $n \times n$ matrix, which has a complexity of $\mathcal{O}(n^2 2^n)$ \cite{ryser1963combinatorial}. This discrepancy highlights a limitation in the current use of tensor networks for large-scale ideal or nearly-ideal systems, despite their success in simulating setups with noise or imperfections \cite{oh2023classical,complexity_TN}.
In this work we formulate an alternative tensor-network-based approach, the operator-basis MPS, to bridge this complexity gap.
{In contrast to the standard approach, where the evolution is represented by MPOs generated by the Hamiltonian in Fock space and subsequently applied to the initial state, our formalism exploits the known input-output relations of the circuit. For an $M$-mode interferometer, these relations are encoded in the matrix $\mathcal{U} = \{u_k^{i}\} \in SU(M)$, with $i, k \in [1,M]$ labeling the input and output ports, respectively. This allows us to directly construct the final state without explicitly performing time propagation, thereby avoiding the primary computational cost of that procedure, which is largely associated with compression.}
To demonstrate the effectiveness of our formalism, we show that our tensor network description of quantum optical circuits can efficiently compute photo-detection amplitudes, which correspond to the permanent of an \( n \times n \) matrix,
\begin{equation}
    \mathrm{Perm}(A) = \sum_{\sigma \in S_n} \prod_{i=1}^n A_{i,\, \sigma(i)}\,,
\end{equation}
where \( \sigma \) ranges over all permutations in the symmetric group \( S_n \). Our approach achieves the optimal classical computational complexity, matching that of Ryser’s algorithm—the best known method for evaluating matrix permanents.
Then we introduce an alternative graphical representation for our MPS, which not only simplifies the overall language but also provides valuable insights, particularly in understanding the scaling of complexity. This approach also suggests a potential strategy for implementation in terms of binary matrices. Finally, we demonstrate that typical imperfections in optical circuits such as partial distinguishability and photon losses can also be treated within our formalism.\\
\section{Operator-basis MPS representation of evolved bosonic ladder operators.}~
A passive linear optical interferometer represented by the unitary $\mathcal{U}=\{u^i_k\}$ acting on $M$ bosonic modes turns the single ladder operator $\Aod{i}$ acting on input mode $i$ into a linear combination $\sum_{k=1}^M u^{i}_k \Aod{k}$ which, in terms of augmented vectors (see Appendix \ref{app:proof_binom}) can be expressed as
\begin{align}
\mathcal{U}(\Aod{i}) & = 
\left( \begin{matrix}  u^{i}_1 \Aod{1} & \mathbb{1} \end{matrix} \right)  \left( \begin{matrix} \mathbb{1} & \mathbb{0} \\u^{i}_2 \Aod{2} & \mathbb{1} \end{matrix}\right)\ldots\left( \begin{matrix} \mathbb{1} & \mathbb{0} \\u^{i}_{M-1} \Aod{M-1} & \mathbb{1} \end{matrix}\right) \notag
\\& \times \left( \begin{matrix} \mathbb{1} \\ u^{i}_M \Aod{M} \end{matrix} \right) 
\,,
\label{eq:matrix_notation}
\end{align}
where, for convenience, we use the same symbol for the channel and its $SU(M)$ representation. Unless otherwise stated, we assume that our evolved operators act on the vacuum state. Considering the local reduced operator basis $\{\hat{a}_{k}^{n \,\dag}\}_{n=0,1}$, the expression above readily maps to the MPO
{
\begin{align}
\mathcal{U}(\Aod{i})&=
\!\!\!\!\!\!\!\
\sum_{\sigma_1,\dots,\sigma_M}\!\!\!
\!\! A^{[1]}_{i\, \sigma_1}  A^{[2]}_{i\, \sigma_2} \ldots\! A^{[M]}_{i\, \sigma_M}
~\bigotimes_{k=1}^M (\hat{a}_{k}^{\dag})^{\sigma_k }\,, 
\label{eq:mpo1}
\end{align}
}
where the tensors $A_i^{[m]\,\sigma_j}$, with $\sigma_j=0,1$ are defined as
\begin{equation}
\begin{aligned}
&A^{[m=1]}_{i\,0} = \left( \begin{matrix} 0 & 1 \end{matrix} \right) , ~ &&A^{[m=1]}_{i\,1} = \left( \begin{matrix} u^{i}_1 & 0 \end{matrix} \right),
\\
&A^{[m\neq 1,M]}_{i\,0} = \left( \begin{matrix} 1 & 0 \\ 0 & 1 \end{matrix} \right), ~ &&A^{[m\neq 1,M]}_{i\,1} = \left( \begin{matrix} 0 & 0 \\ u^{i}_m & 0 \end{matrix} \right), 
\\
&A^{[m=M]}_{i\,0} = \left( \begin{matrix} 1 \\ 0 \end{matrix} \right) , ~ &&A^{[m=M]}_{i\,1} = \left( \begin{matrix} 0 \\ u^{i}_M \end{matrix} \right),
\label{eq:matrices}
\end{aligned}
\end{equation}
where $1 \le m \le M${, and generally $\sigma_1,\dots,\sigma_M \in \{0,1\}$. However, we note} that for $m\neq 1,M$ the $A^{[m]}_{i\,1}$-matrices are nilpotent with index $2$. As a consequence, the sum in the equation above can be restricted to terms satisfying $\sum_{k=1}^M \sigma_k =1$. 

The bond dimension of the MPO in \eq\eqref{eq:mpo1} is upper bounded by $D=2$ for any passive linear optics network, which is consistent with the result in \cite{Cilluffo2024boson} and \cite{PhysRevLett.102.057202,clark2010exact} for local fermionic operators. When considering a product of $n$ ladder operators
$\Aod{\bar{i}}\equiv\Aod{i_1} \Aod{i_2} \ldots \Aod{i_n}$, evolving as the tensor product of each $\mathcal{U}(\Aod{i_j})$,
\eq\eqref{eq:mpo1} generalizes to
\begin{align}
\mathcal{U}(\Aod{\bar{i}})= \!\!\!\!\!\!\!\sum_{\sum_{m=1}^M n_m =n} 
\!\!\!\!\!\!\!  \mathcal{A}_{\bar{i}}^{[1]\, n_1} \mathcal{A}_{\bar{i}}^{[2]\, n_2}\ldots \mathcal{A}_{\bar{i}}^{[M]\, n_M} ~\bigotimes_{k=1}^M (\hat{a}_{k}^{\dag})^{n_k }\,,
\label{eq:mpo2}
\end{align}
where
{\begin{align}
\sqrt{n_m!} \mathcal{A}_{\bar{i}}^{[m]\, n_m}  = \!\!\!  \sum_{\sum_{k=1}^{ n} \sigma^{(m)}_k = n_m} \!\!
A^{[m]}_{i_1\,\sigma^{(m)}_1} \otimes \ldots \otimes A^{[m]}_{i_n \,\sigma^{(m)}_n}\,.
\label{eq:expansion}
\end{align}
}
Probability amplitudes for specific Fock space components can be retrieved by computing the products of the $\mathcal{A}$ matrices in \eq\eqref{eq:mpo2}.
{
We conclude this part by noticing that, while these objects are formally MPOs, in operator space their tensor structure along the chain of modes, with one physical index per local operator, closely resembles an MPS. Importantly, these OBMPS are always meant to be applied to the vacuum; in this sense, they encode all the information about the corresponding Fock-state amplitudes. In this structural sense, we define them as operator-basis MPS (OBMPS).
}
\section{Graphical interpretation.}
The expression in \eq\eqref{eq:expansion} is well-suited for a graph-based interpretation, which provides a more straightforward approach to the calculations.
We define
\begin{equation}
\begin{aligned}
    \ws\ \: &:=A^{[m]}_{i\,0} = \mathbb{1}_2  ~~~~~~~ &&\gs:=m \left( \begin{matrix} 0 & 0 \\ 1 & 0 \end{matrix} \right)\,
\end{aligned}
\end{equation}
{ where $m$ is a complex number. We also define}
the two operations
\begin{equation}
\begin{aligned}
    \gws ~ \gws  &:= X_i^{[m]} \otimes X_j^{[m]} ~ &&\begin{matrix} \gws \\[-1pt] \gws  \end{matrix} := X_i^{[m]} \times X_i^{[m+1]}\,
\end{aligned}
\end{equation}
{where the generic tensor $X:=\gws$ can correspond either to a green or a whit square.}
Note that the nilpotence of the A-matrices translates as
$ \begin{matrix} \gs \\[-1pt] \gs  \end{matrix} = 0\,.$
The \eq\eqref{eq:expansion} translates into
\begin{align}
\mathcal{A}_{\bar{i}}^{[j]\, n_j} = [ \, \underbrace{\gs ~ \gs \ldots \gs}_{n_j} ~ \underbrace{\ws \ldots \ws}_{n-n_j}\, ],
\end{align}
{ where the square brackets denote the sum over all $\binom{n}{n_j}$ possible placements of $n_j$ green squares ($\gs$) and $n-n_j$ white squares ($\ws$). 
}
We will refer to this type of graph and to each arrangement as \textit{lines} and \textit{components} respectively.
Thus, from \eq\eqref{eq:mpo2}, the amplitude $\alpha_{\bar{n}}$ corresponding to a component $\Aod{\bar{j}}|0 \rangle = | \bar{n} \ra$ of the final state is turned into a grid of $M$ lines stacked vertically
\begin{align}
\alpha_{\bar{n}}= \la 0|\Ao{\bar{j}} ~ \mathcal{U}(\Aod{\bar{i}})|0\ra = ~ 
 \underbrace{\begin{matrix}
[\,\gs ... \gs ~ \ws ...  \ws ~ \ws ...  \ws ~ \ws ...  \ws \,]\\
[\, \ws ... \ws ~ \gs ... \gs ~ \ws ...  \ws 
~ \ws ...  \ws \,]\\
...  \\
[\,\ws ... \ws ~ \ws ... \ws ~ \ws ...  \ws ~ \gs ...  \gs \,] \\
    \end{matrix} }_{n = |\bar{n}|} \,
\, .
\end{align}
We will refer to the contraction of two lines, i.e.~to the product of two $\mathcal{A}$ matrices, as \textit{merging}. In terms of graphs
\begin{align}
\mathcal{A}_{\bar{i}}^{[j]\, n_j} \mathcal{A}_{\bar{i}}^{[j']\, n_{j'}} = [\,
\underbrace{\gs ~ \gs \ldots \gs}_{n_j+n_{j'}} ~ \underbrace{\ws \ldots \ws}_{n-n_j-n_{j'}} ]\,.
\end{align}
The elementary properties of these operations along with proofs and examples will be presented in detail in Appendix \ref{app:proof_binom}.
\section{Inverse change of basis.}~
While we focus on computing probability amplitudes, the formalism is not limited to this. The operator-based construction can in principle be used for general-purpose state manipulation, beyond computing expectation values. For this reason we complete the set of basic rules of the formalism by defining the transformation of bosonic operator-basis MPSs into standard MPSs.
We leverage two key insights: the Kronecker product of multiple $A$ matrices results in a sparse matrix and the positions of the elements of $\mathcal{A}$ corresponding to different powers of $\Aod{}$ follow regular patterns.
Regarding the latter, as demonstrated e.g.~in \cite{ZHOU2023112962}, the multiple tensor products of $2 \times 2$ triangular matrices generate a $2^n \times 2^n$ matrix whose elements are distributed according to a fractal structure known as the \textit{Sierpi\'nski triangle}.
In our case, we observe the same phenomenon for the repeated tensor products of the elements in our decomposition. This is illustrated below for a 3-photon tensor
\begin{align}
&\left( \begin{matrix} \mathbb{1} & \mathbb{0} \\u^i_m \Aod{} & \mathbb{1} \end{matrix}\right) \otimes \left( \begin{matrix} \mathbb{1} & \mathbb{0} \\u^j_m \Aod{} & \mathbb{1} \end{matrix}\right) \otimes \left( \begin{matrix} \mathbb{1} & \mathbb{0} \\u^k_m \Aod{} & \mathbb{1} \end{matrix}\right) 
\notag\\&= 
\left( \everymath{\scriptstyle}
\begin{matrix} 
\mathbb{1} & \mathbb{0} & \mathbb{0} & \mathbb{0} & \mathbb{0} & \mathbb{0} & \mathbb{0} & \mathbb{0} 
\\
u^k_m \Aod{} & \mathbb{1} &\mathbb{0} & \mathbb{0} & \mathbb{0} & \mathbb{0} & \mathbb{0} & \mathbb{0}  
\\
u^j_m \Aod{} & \mathbb{0} &\mathbb{1} & \mathbb{0} & \mathbb{0} & \mathbb{0} & \mathbb{0} & \mathbb{0}  
\\
u^{j k}_m \hat{a}^{\dag\,2} & u^j_m \Aod{} & u^k_m \Aod{} & \mathbb{1} & \mathbb{0} & \mathbb{0} & \mathbb{0} & \mathbb{0} 
\\
u^i_m \Aod{} & \mathbb{0}  & \mathbb{0} & \mathbb{0} & \mathbb{1} & \mathbb{0} & \mathbb{0} & \mathbb{0} 
\\
u^{i k}_m \hat{a}^{\dag\,2} & u^i_m \Aod{} & \mathbb{0} & \mathbb{0}   & u^k_m \Aod{}  & \mathbb{1} & \mathbb{0} & \mathbb{0} 
\\
u^{i j}_m \hat{a}^{\dag\,2} & \mathbb{0} & u^i_m \Aod{} & \mathbb{0} & u^j_m \Aod{} & \mathbb{0} & \mathbb{1} & \mathbb{0} 
\\
u^{ijk}_m \hat{a}^{\dag\,3} & u^{ij}_m \hat{a}^{\dag\,2} & u^{ik}_m \hat{a}^{\dag\,2} & u^i_m \Aod{} & u^{j k}_m \hat{a}^{\dag\,2} & u^j_m \Aod{} & u^k_m \Aod{} & \mathbb{1} 
\end{matrix}\right)
\,,
\label{eq:example_syer}
\end{align}
where we used the shorthand notation $u^{ijk}_m:=u^{i}_m u^{j}_m u^{k}_m$.
Tensors like the one above
can be turned into a standard Fock-basis tensor network by replacing each component with the corresponding $d\times d$ matrix, being $d$ the local dimension cutoff.
However, thanks to the regular and sparse structure of the Sierpiński triangle, each and every component can be computed independently of all others. This enables the encoding of components with a specific photon number $n_j < n$ of the single-mode tensor $\mathcal{A}^{[j]\, n_j}_{\bar{i}}$ without generating and storing all possible components. Furthermore, the logical operations required to produce the fractal structure can be executed efficiently and in parallel using binary strings. The Sierpi\'nski triangle corresponding to the matrices $\mathcal{A}^{[m]\,n}_{\bar{i}}$ for a given number of photons $n$ in the target interferometer can be generated recursively from the set of the lexicographically ordered subsets $[(),u_m^{i_1} \Aod{m},u_m^{i_2} \Aod{}, u_m^{i_1} u^{i_2} ({\Aod{m}})\,^2, \ldots, u_m^{\bar{i}} ({\Aod{m}})\,^n]$ with non-empty intersection, as depicted in Fig. \ref{fig:fig3} for a $3$-photon scenario (see Appendix \ref{app:sy} for details).
Therefore we are able to generate exclusively the required components once $n_j$ is fixed, offering significant advantages in memory usage and computational efficiency.
\begin{figure}
\centering
\includegraphics[scale=0.3,angle=0]{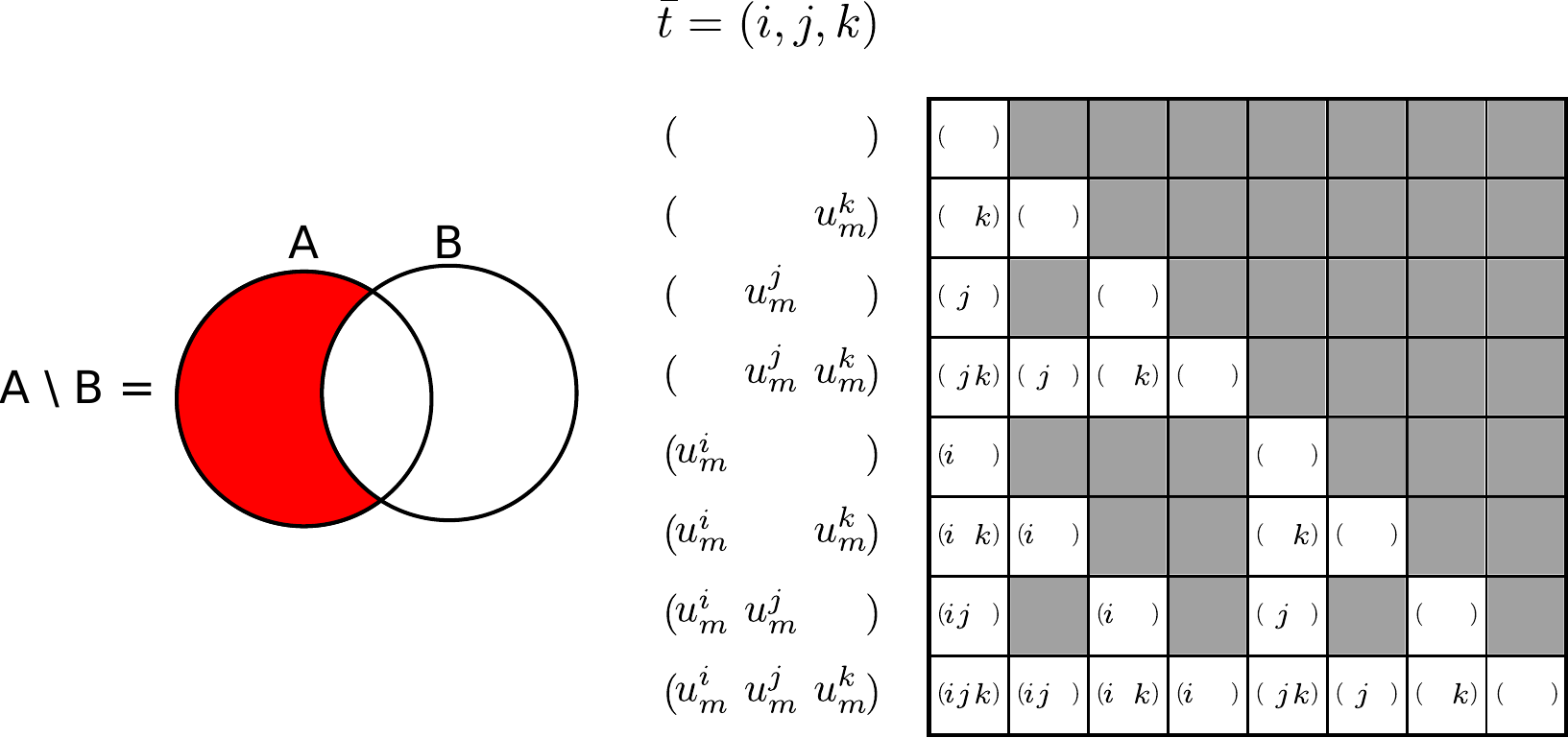}	
\caption{Left side: the relative complement or set difference between the sets $A$ and $B$, is defined as $A \backslash B=\{x \in A : x\notin B\}$. On the right: example of generation of the Sierpi\'nski triangle structure for the tensor in \eq\eqref{eq:example_syer}. The subsets of $(u_m^i\,u_m^j\,u_m^k)$ are arranged in lexicographic order, $()$ represents the empty set and grey squares correspond to the cases in which $A \cap B = ()$ (see Appendix \ref{app:sy} for details).
} 
\label{fig:fig3}
\end{figure}
\\
\textit{Composition of channels.}~
We consider the representation of the composition of unitary channels {$\mathcal{U}$ and $\mathcal{V}$, denoted by $\mathcal{V}\circ \mathcal{U}$,} within the bosonic-operator-basis-MPS formalism.
We will restrict to single-photon states, as the many-photon case is obtained through tensor products.
In this scenario, the input-output matrix 
that represents the evolution is $\mathcal{U} \cdot \mathcal{V}$.
The state of a single photon at the input port $i$ evolved through the composition of unitary channels $\mathcal{V}\circ\mathcal{U}$ reads
\begin{align}
&\mathcal{V}\circ \mathcal{U}\left( \Aod{i}
\right)= \notag
\\&=
 \left(\begin{matrix} u^i_1 \sum\limits_{j=1}^{ M} v^1_j \Aod{j} & \mathbb{1} \end{matrix}\right) \prod_{k=2}^{M-1}
 \left(\begin{matrix} \mathbb{1} & \mathbb{0} \\  u^i_k \sum\limits_{j=1}^{ M} v^k_j \Aod{j} & \mathbb{1} \end{matrix}\right)  
 \left(\begin{matrix} \mathbb{1} \\  u^i_{ M} \sum\limits_{j=1}^{ M} v^{ M}_j \Aod{j}   \end{matrix}\right) 
 \notag\\&
= \left(\begin{matrix} \sum\limits_{j=1}^{ M} u^i_j v^j_1 \Aod{1} & \mathbb{1} \end{matrix}\right) 
 \prod_{k=2}^{M-1} \left(\begin{matrix} \mathbb{1} & \mathbb{0} \\  \sum\limits_{j=1}^{ M} u^i_j v^j_k \Aod{k} & \mathbb{1} \end{matrix}\right)
 \left(\begin{matrix} \mathbb{1} \\ \sum\limits_{j=1}^{ M} u^i_j v^j_{ M} \Aod{{ M}}   \end{matrix}\right) \,,
\label{eq:composite0}
\end{align}

where $\mathcal{V} = \{v^i_j\}$ and we used the commutative property (see Appendix \ref{app:proof_binom}).
As expected, the coefficients in the above equation are the sums of the row elements from the input-output matrix of the combined transformation. Notably, for each product of coefficients, the first upper index on the left, $i$, denotes the input port, while the final lower index on the right represents the output port through which the photon exits the interferometer. Therefore, as a consistency check, if $\mathcal{U} = \mathcal{V}^{-1}$, all coefficients will vanish except for $\sum_{j=1}^{n} u^i_j v^j_i$, and the state simplifies to the single ladder operator at $i$. Equation \eqref{eq:composite0} can be readily generalized to scenarios with multiple channels using the following Lemma, which directly follows from the commutative property (see Appendix \ref{app:proof_binom} for details):
\begin{lemma}
Let ${u^{(k)\,i}_j}$ be the elements of a generic $k$th transformation $\mathcal{U}^{(k)}$ acting on the ladder operator $\Aod{i_1}$, and $T_{i_2,i_3,\ldots, m}^{i_1,i_2,\ldots,i_n}=u^{(1)\,i_1}_{i_2} u^{(2)\,i_2}_{i_3}\ldots u^{(n)\,i_n}_{m} \Aod{m}$, with $m \in [1,M]$, the non-identity element of the $m$th matrix in the expansion of $\bigcirc_{k=1}^n \mathcal{U}^{(k)}$. Then, $\forall m$, $\mathcal{U}^{(n+1)} \circ\bigcirc_{k=1}^n \mathcal{U}^{(k)}$ corresponds to the following transformation on the non-identity elements
\begin{align}
T_{i_2,i_3,\ldots,i_n, m}^{i_1,i_2,i_3\ldots,i_n} \rightarrow 
T_{i_2,i_3,\ldots,i_n,i_{n+1},m }^{i_1,i_2,i_3\ldots,i_n,i_{n+1}}\,.
\end{align}
\end{lemma}
This concludes the basic rules of our formalism. We now move on to benchmark it by calculating the probability amplitude for a specific photon-counting outcome, starting from a given input state. The square modulus of this amplitude corresponds to the coincidence rate in a multiphoton interference experiment.
\section{Multiphoton detection probability amplitudes.}
 For the sake of argument we consider the case $n_i=1$ for $i\le n$ and $0$ otherwise \cite{note2}. This is known to correspond to the permanent of a $n\times n$ complex matrix \cite{aaronson2011computational}.
The diagrammatic representation of the amplitude $\alpha_{\bar{n}}$ reads
\begin{align}
\alpha_{\bar{n}} = ~ \left.
\begin{matrix}
[\, \gs ~ \ws ~ \ws \ldots \ws \,]\\
[\, \ws ~ \gs ~ \ws \ldots \ws \,]\\
\ldots \\
[\, \ws ~ \ws ~ \ws  \ldots \gs \,]\\
\end{matrix} ~~\right\} n\, .
\label{eq:graph_perm}
\end{align}
There is not a unique contraction strategy. 
The direct computation of all scalar products followed by tensor products results in the same computational scaling as using the definition of permanents directly (see Appendix \ref{app:naive}). We show here how we can obtain a computational advantage
over the standard tensor network approach by working with the diagrams and the given rules.
First, we analyze the computational cost of merging the first two lines in \eq\eqref{eq:graph_perm}. 
\begin{align}
\begin{matrix}
[\, \gs ~ \ws ~ \ws \ldots \ws \,]\\
[\, \ws ~ \gs ~ \ws \ldots \ws \,]\\
\end{matrix} ~ = [\, \gs ~ \gs ~ \ws \ldots \ws  \,]\,.
\end{align}
Due to the nilpotence of the $A^{[m]1}$-matrices, each of the $n$ component of the first line can be merged with $n-1$ components of the second one.
In Appendix \ref{app:proof_binom} we show that the number of operations necessary to perform these operation is $\mathcal{O}(n)$. Thus the resulting computational cost of merging the first two lines is $n \binom{n}{1} \binom{n-1}{1}$.
The same arguments hold for the next contraction, 
\begin{align}
\begin{matrix}
[\, \gs ~ \gs ~ \ws ~\ws \ldots \ws \,]\\
[\, \ws ~ \ws ~ \gs ~\ws \ldots \ws \,]\\
\end{matrix} ~ = [\, \gs ~ \gs ~ \gs ~\ws \ldots \ws \,] \,,
\end{align}
with the difference that this time the nilpotence of the matrices implies that for a fixed component in the first line we have $\binom{n}{2}$ different components in the subsequent line with non-zero contribution. This gives $n \binom{n}{2} \binom{n-2}{1}$ operations .
By proceeding in the same way for the rest of the operations we find that the computational cost $c$ is given by
\begin{align}
c = n \cdot \sum_{i=1}^{n-1} \binom{n}{i} \binom{n-i}{1}
= n^2(2^{n-1}-1) = \mathcal{O}(n^2 2^{n})
\label{eq:cost_permanent}
\end{align}
This matches the computational complexity of computing the permanent of a $n\times n$ matrix using Ryser's algorithm. 
Thus the bosonic-operator-basis MPS representation provides a computationally optimal description of linear interferometers followed by photon-counting measurements. Furthermore, relaxing the condition $n_i=1$ reduces the number of merging operations required to compute the amplitude, consistent with the reduction in complexity observed for Boson Sampling when multiple photons occupy the same input mode \cite{aaronson2012generalizing}.
\section{Photonic circuits with imperfections}~
Partial distinguishability of photons introduces important distortions in quantum interference effects and constitutes one of the main sources of imperfection in boson-sampling experiments.
In order to introduce this feature, we consider now an internal and, for the sake of clarity, dichotomic degree of freedom for each photon, namely
\begin{align}
\Aod{}\rightarrow \sqrt{\eta}\,\Aod{} + \sqrt{1-\eta}\,\hat{a}^{\perp\,\dag}\,,
\end{align}
where $\eta\in [0,1]$ encodes the partial distinguishability. 
The components generated by the two ladder operators are mutually orthogonal and do not interact within the interferometer. Consequently, the formalism described earlier can be easily extended by using direct sums of orthogonal subspaces. Thus, based on \eq\eqref{eq:matrices}, we introduce the operator basis for distinguishable photons
\begin{align}
&B^{[m]}_{i\,0} = A^{[m]}_{i\,0} \oplus A^{[m]}_{i\,0}; ~  
\\&
B^{[m]}_{i\,1} = \sqrt{\eta_i} \, A^{[m]}_{i\,1} \oplus \sqrt{1-\eta_i} \, A^{[m]}_{i\,1}\, \label{eq:matrices_dist}
\end{align}
where the distinguishability coefficient refers to the photon entering through the $i$-th port. The interferometer will act independently on both the sides of the direct sum.
The distributivity of the matrix multiplication over the direct sum ensures that the bond dimension does not change.
In the limiting case of many photons being perfectly distinguishable (e.g., each with a unique frequency), the direct sum defined above introduces terms from additional subspaces, effectively summing over contributions from these subspaces. Consequently, the computation of a multiphoton amplitude simplifies to a sum of single-photon contributions, significantly reducing computational complexity. In a less extreme scenario where groups of photons are partially distinguishable, this leads to the sum of smaller permanents, which is consistent with the findings in \cite{PhysRevLett.120.220502}.
Analogously, photon losses can be integrated with minimal modifications { by considering the enlarged matrix  $\tilde{\mathcal{U}}$ which includes additional sink modes representing the environment.}
{ In the scenario of uniform losses, the linear transformation is $(\sqrt{1-\epsilon} ~\mathcal{U} ~~ \sqrt{\epsilon} \mathbb{1}_M)$.
For each input photon, any loss process corresponds to the creation of an excitation in one of these extra modes.}
As a consequence the \eq\eqref{eq:matrix_notation} is modified as follows
\begin{align}
\tilde{\mathcal{U}}(\Aod{i})
&= 
\left( \begin{matrix} \tilde{u}^{i}_{1} \Aod{1} & \mathbb{1} \end{matrix} \right)  \left( \begin{matrix} \mathbb{1} & \mathbb{0} \\\tilde{u}^{i}_2 \Aod{2} & \mathbb{1} \end{matrix}\right) 
\ldots\left( \begin{matrix} \mathbb{1} \\ { \sqrt{\epsilon} \hat{a}_{M+i}^\dag } \end{matrix}\right)
\,.
\end{align}
{ where $\tilde{u}^{i}_j=\sqrt{1-\epsilon} u^{i}_j$.}
\begin{figure}
\centering
\includegraphics[scale=0.7,angle=0]{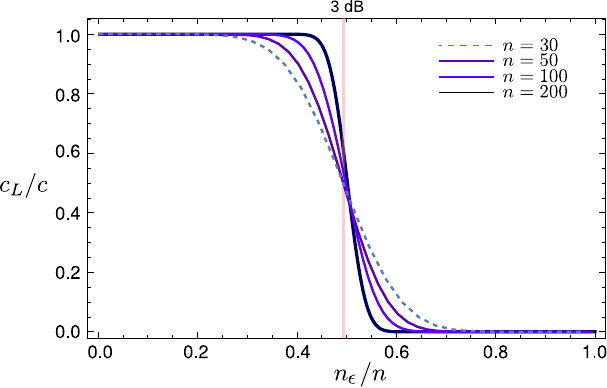}	
\caption{Ratio between the computational cost of the permanent $c$ (\eq\eqref{eq:cost_permanent}) and $c_L$ as a function of the fraction of photon lost, $n_{\epsilon}/n$, for different numbers of input photons $n$. The drop in complexity corresponds to $50\% \sim 3\,{\rm dB}$ losses.}
\label{fig:fig4}
\end{figure}
The formalism outlined in the previous section can now be applied directly without significant modification.
The probability amplitude $\alpha_{\bar{n}}$, given the number of input photons $n\ge n_{\rm out}=\sum_{i=1}^M n_i$ reads
\begin{align}
( \mathcal{A}_{\bar{i}}^{[1]\, n_1} \mathcal{A}_{\bar{i}}^{[2]\, n_2} \!\!...\, \mathcal{A}_{\bar{i}}^{[M]\, n_M} ) ( \mathcal{A}^{[M+1]\,n_{M+1}}_{\bar{i}} \!\!\!\!...\, \mathcal{A}^{[2M]\,n_{2M}}_{\bar{i}}) \,,
\label{eq:lossy_prob}
\end{align}
{with
$\mathcal{A}^{[M+k]\,n_{M+k}}_{\bar{i}}$ expressed as in Eq.~\eqref{eq:expansion} with the $A$-matrices like in Eq.~\eqref{eq:matrices} but with all coefficients $u$ replaced by $\sqrt{\epsilon}$.
Since the coefficients in the lossy sector are uniform, any contraction involving a fixed number of lost photons contributes the same overall factor, independent of how the losses are distributed among the sink modes. This common prefactor can therefore be computed separately, and in the graphical representation the entire loss contribution reduces to a single effective line that accounts for all lost photons:
}
\begin{align}
\begin{matrix}
n_{out} \left\{\begin{matrix}
[\,\gws ~ \gws ~ \ldots ~ \gws ~ \gws \ldots \gws \,]\\
[\,\gws ~ \gws ~ \ldots ~ \gws ~ \gws \ldots \gws \,]\\
\ldots\\
[\,\gws ~ \gws ~ \ldots ~ \gws ~ \gws \ldots \gws \,]\\
\end{matrix}\right. \\
\,~~~~~~~~\,\underbrace{\!\!\tys [\, \gs ~ \gs ~ \ldots ~ \gs}_{n_{\epsilon}} \tys\tys\tys\tys \ws \ldots \tys\tys \ws \,]
\end{matrix} 
\,,
\end{align}
{ where $n_\epsilon=\sum_{i=M}^{2M} n_i$.}
Note that each component resulting from the merging of the first $n_{out}$ lines determines a unique component in the last line. 
As a consequence, computing the probabilities conditioned to a specific amount $n_{\epsilon}$ of photons lost, reduces to the same procedure previously described, with a number of arithmetic operations scaling as $c_L = n^2 \sum_{i=1}^{n-n_{M+1}} \binom{n-1}{i} < n^2 2^n$.
In Fig.~\ref{fig:fig4} we report the ratio $c_L/c$ as a function of the number of photon lost. A distinct threshold emerges at $3~{\rm dB}$ of losses, which aligns with the best state-of-the-art experiments \cite{oh2024classical}. We conclude this section by addressing dephasing. In the pure-state framework adopted here, dephasing can be modeled by introducing random phase shifts in the matrix representation of the channel. In photon counting experiments, this results in expectation values that depend on these phases, which can be averaged a posteriori. This process requires an external routine, such as Monte Carlo integration, which is fully parallelizable as long as the phase shifts are independent. However, handling arbitrary defects would be more effectively approached using a density-matrix formalism, which we leave for future exploration.
\section{Discussion}~
{  
We leverage the known input–output relations of linear optical circuit components to directly construct, through an MPS-like representation, the exact operator expression defining the output of a linear optical network, thereby avoiding explicit time propagation.}
The graphical representation of our tensors simplifies the identification of optimal contraction strategies and allows for a precise evaluation of the computational scaling involved in calculating probability amplitudes and expectation values of observables expressed as products of bosonic creation and annihilation operators.
Specifically, in the context of Boson Sampling experiments, this approach allows to compute permanents of complex matrices with the same scaling as the best-known algorithms. In contrast, until now, the use of tensor networks has been limited to small, idealized setups or larger interferometers plagued by a high degree of defects, which degrade correlations and undermine the claim of quantum advantage \cite{oh2023classical,complexity_TN}.
An operator-basis MPS can also be efficiently converted into a standard MPO and applied to a state in the Fock basis with minimal overhead. Notably, translating a bosonic operator-basis MPS into a standard MPS via \eq\eqref{eq:matrix_notation} involves $\mathcal{O}(2^{2n})$ operations, identifying the reason for the
complexity gap between our representation and conventional tensor network methods. Moreover, our approach suggests a potential advantage by leveraging the fractal structure of the $\mathcal{A}$-matrices, offering a more efficient alternative to the standard translation.
We have shown how the proposed formalism can be used to describe important sources of
defects occurring in quantum optical experiments, namely partial distinguishability, photon losses, {in particular uniform losses,} and dephasing.
In particular, we can establish thresholds on the level of imperfections 
an experiment can tolerate while preserving the quantum advantage regime. 
 Provided that the effects of such defects on the dynamical map governing
 the system's evolution are accurately characterized and represented in the input-output relations of the channel, our approach is well-suited to incorporate them.
Finally, employing this formalism within the Heisenberg picture offers substantial advantages for simulating observables such as photon-counting moments. This is especially valuable when dealing with highly complex input states (e.g., those with high photon numbers, strong correlations, or non-Gaussian features) or when a transition to the density matrix formalism becomes necessary.
Moreover, the same framework can be extended to describe coherent states (displaced vacuum) and squeezed states on specific modes by incorporating displacement and squeezing operators as transformations of the local $A$-matrices.
\section*{Acknowledgements}
We acknowledge T. Lacroix, G. W\'ojtowicz and G. Magnifico for fruitful discussions.
This research was supported by the BMBF project PhoQuant (Grant No. 13N16110).

\bibliography{biblio}

\clearpage

\appendix

\section{Augmented vectors algebra and combination rules for graphs}
\label{app:proof_binom}

In this appendix, we outline the fundamental properties of augmented vectors and establish the combination rules for $\mathcal{A}$-matrices and their associated graphs.

The matrix representation in \eq\eqref{eq:matrix_notation} corresponds to mapping a sum of operators onto an equivalent product of augmented vectors according to the following

\begin{definition}{Vectorized sum}
\begin{align}
\sum_{i=1}^{n} \mathbb{x}_i= \left(\begin{matrix} \mathbb{x}_1 & \mathbb{1} \end{matrix}\right)
\left(\begin{matrix} \mathbb{1} & \mathbb{0} \\ \mathbb{x}_2 & \mathbb{1} \end{matrix}\right) \ldots \left(\begin{matrix} \mathbb{1} & \mathbb{0} \\ \mathbb{x}_{n-1} & \mathbb{1} \end{matrix}\right)\left(\begin{matrix} \mathbb{1} \\ \mathbb{x}_n  \end{matrix}\right)
\end{align}
where $\mathbb{x}_i$ are elements of a generic vector space over $\mathbb{C}$ and $\mathbb{1}$ is the neutral element of multiplication.
\end{definition}

The distributive property of multiplication and commutative property of the sum translate straightforwardly:

\begin{property}{Distributive property of the product by a scalar}
\begin{align}
\lambda&\left(\begin{matrix} \mathbb{x}_1 & \mathbb{1} \end{matrix}\right)
\left(\begin{matrix} \mathbb{1} & \mathbb{0} \\ \mathbb{x}_2 & \mathbb{1} \end{matrix}\right) \ldots \left(\begin{matrix} \mathbb{1} & \mathbb{0} \\ \mathbb{x}_{n-1} & \mathbb{1} \end{matrix}\right)\left(\begin{matrix} \mathbb{1} \\ \mathbb{x}_n  \end{matrix}\right) \nonumber\\
& \;\;\;\; = 
\left(\begin{matrix} \lambda \mathbb{x}_1 & \mathbb{1} \end{matrix}\right)
\left(\begin{matrix} \mathbb{1} & \mathbb{0} \\ \lambda \mathbb{x}_2 & \mathbb{1} \end{matrix}\right) \ldots \left(\begin{matrix} \mathbb{1} & \mathbb{0} \\ \lambda\mathbb{x}_{n-1} & \mathbb{1} \end{matrix}\right)\left(\begin{matrix} \mathbb{1} \\ \lambda \mathbb{x}_n  \end{matrix}\right) 
\end{align}
\end{property}

\begin{property}{Commutative property of the sum}
\label{prop:comm}
\begin{align}
&\left(\begin{matrix} \sum_{i=1}^{n} \mathbb{x}^{(1)}_i & \mathbb{1} \end{matrix}\right)
\left(\begin{matrix} \mathbb{1} & \mathbb{0} \\  \sum_{i=1}^{n} \mathbb{x}^{(2)}_i & \mathbb{1} \end{matrix}\right) \ldots \left(\begin{matrix} \mathbb{1} \\ \sum_{i=1}^{n} \mathbb{x}^{(M)}_i  \end{matrix}\right) 
\notag\\&= 
\left(\begin{matrix} \sum_{i=1}^{M} \mathbb{x}^{(i)}_1 & \mathbb{1} \end{matrix}\right)
\left(\begin{matrix} \mathbb{1} & \mathbb{0} \\  \sum_{i=1}^{M} \mathbb{x}^{(i)}_2 & \mathbb{1} \end{matrix}\right) \ldots \left(\begin{matrix} \mathbb{1} \\ \sum_{i=1}^{M} \mathbb{x}^{(i)}_n  \end{matrix}\right) 
\end{align}
\end{property}

\begin{property}
    All the components of two lines with at least one green square in the same position return $0$ in merging, i.e.
   \begin{equation}
        \begin{matrix}
            \gws ~ \gs \ldots \gws \\
            \gws ~ \gs \ldots \gws \\
    \end{matrix} =0
    \end{equation}
\end{property}
\textit{Proof.}
The distributive property of the matrix product over the tensor product implies that 
\begin{align}
\mathcal{A}_{\bar{i}}^{[j]\, n_j} \mathcal{A}_{\bar{i}}^{[j']\, n_{j'}} \!\!\! &= \!\!\!\!\!\!\!\!\!  \sum_{\sum_{k=1}^M \sigma^{(j)}_k = n_j} \!\!\!\!\!\!\!\!\!
A^{[j]}_{i_1\,\sigma^{(j)}_1}  A^{[j']}_{i_1\,\sigma^{(j')}_1} 
\otimes \ldots \otimes A^{[j]}_{i_n \,\sigma^{(j)}_n} A^{[j']}_{i_n \,\sigma^{(j')}_n}\,.
\end{align}
Due to the nilpotence of the A-matrices, $A^{[j]}_{i_m\,\sigma^{(j)}_m}\times A^{[j']}_{i_m\,\sigma^{(j')}_m}$ returns a null vector if $\sigma^{(j)}_m = \sigma^{(j')}_m = 1$,  which corresponds to a graph with two green squares vertically aligned. As a consequence, the product vanishes.
\QED

\begin{property}
    Merging two lines with $n$ and $n'$ green squares respectively results in a line with $n+n'$ green squares.
    \label{prop:merg}
\end{property}

{
\begin{example}
To illustrate the property, consider the following component contractions ~  
\begin{align}
\begin{matrix}
\gs ~ \ws ~ \gs ~\ws ~\ws \ldots \ws~~~~~~
\ws ~ \ws ~ \gs ~\gs ~\ws \ldots \ws\\
\ws ~ \ws ~ \ws ~\gs ~\ws \ldots \ws~~~,~~\gs ~ \ws ~ \ws ~\ws ~\ws \ldots \ws\\
\\
\gs ~ \ws ~ \ws ~\gs ~\ws \ldots \ws\\
\ws ~ \ws ~ \gs ~\ws ~\ws \ldots \ws\\
\end{matrix} \,,
\end{align}
which result in three components with the same shape:
\begin{align}
\begin{matrix}
\gs ~ \ws ~ \gs ~\gs ~\ws \ldots \ws\
\end{matrix} \,
\label{eq:line_ex}
\end{align}
The sum of these contributions gives 
\begin{align} 
\sum_{i=1}^3 u_i v_i  w_i \left( \begin{matrix} 0 & 0 \\ 1 & 0 \end{matrix}\right)\otimes \mathbb{1}_2 \otimes 
\left( \begin{matrix} 0 & 0 \\1 & 0 \end{matrix}\right) \otimes \left( \begin{matrix} 0 & 0 \\ 1& 0 \end{matrix}\right) \otimes \mathcal{I}
\,,
\end{align}
where $u_i, v_i, w_i$ are the nonzero matrix elements for the three components and $\mathcal{I}$ denotes the tensor product of identities over the remaining squares.
Therefore the result of the three contractions can be combined into a single scalar factor multiplying a single tensor product of 2-tensors.
This sum can always be expressed as a product 
$ \sum_{i=1}^3 u_i v_i w_i= x_1 x_2 x_3$, 
leading to
\begin{align} 
\left( \begin{matrix} 0 & 0 \\x_1 & 0 \end{matrix}\right)\otimes \mathbb{1}_2 \otimes 
\left( \begin{matrix} 0 & 0 \\x_2 & 0 \end{matrix}\right) \otimes \left( \begin{matrix} 0 & 0 \\x_3 & 0 \end{matrix}\right)\otimes \mathcal{I} 
\,.
\end{align}
\end{example}
Thus, in practice, one fixes the pattern of the component resulting from the contraction, collects the corresponding components and coefficients, and then reshapes the result as a single tensor product.
}

The proof of this property { in the general case} requires some technical tools.
We show first that the product of two $\mathcal{A}^{[j]\,n_j}_{\bar{i}} \mathcal{A}^{[j+1]\,n_{j+1}}_{\bar{i}} $ returns $\binom{n}{n_j+n_{j+1}}$ objects.
\begin{property}{}
\label{prop:tp}
Let $\mathbf{s}$ be a binary vector of $n$ elements. Then the non-zero elements of the matrix resulting from the tensor product
\begin{align}
\bigotimes_{k=1}^n A_{i_k\, \sigma_k^{(m)}}^{[m]} \Big|_{\{\sigma_k^{(m)}\}=\mathbf{s}}
\label{eq:tens_prod_res}
\end{align}
take the form 
\begin{align}
\prod_{\{k: s_k = 1 \}} \!\!\!\! u_m^{i_k}\,.
\label{eq:prod_res2}
\end{align}
\end{property}
The following Lemma corresponds to the Property \ref{prop:merg}:
\begin{lemma}{}
\label{lem:2}
Let $\mathbf{r}$, $\mathbf{s}^{(m)}$ and $\mathbf{s}^{(m+1)}$ be binary vectors of $n$ elements and norm $n_m$ and $n_{m+1}$ respectively. Thus for any $\mathbf{r}$
\begin{align}
&\sum_{\{\mathbf{\Omega}\}} \bigotimes_{k=1}^n A_{i_k\, \sigma_k^{(m)}}^{[m]} \Big|_{\{\sigma_k^{(m)}\}=\mathbf{s}^{(m)}} \times \bigotimes_{k=1}^n A_{i_k\, \sigma_k^{(m+1)}}^{[m+1]} \Big|_{\{\sigma_k^{(m+1)}\}=\mathbf{s}^{(m+1)}} \notag
\\&=
\bigotimes_{k=1}^n \tilde{A}_{i_k\, \tilde{\sigma}_k^{(m)}}^{[m+1]}  \Big|_{\{\tilde{\sigma}_k^{(m)}\}=\mathbf{r}} 
\label{eq:l2}
\end{align}
with
\begin{align}
\mathbf{\Omega} 
= 
\left\{
 (\mathbf{s}^{(m)} , \mathbf{s}^{(m+1)} ) :  \mathbf{s}^{(m)} + \mathbf{s}^{(m+1)} = \mathbf{r}
\right\}
\end{align}
and 
\begin{align}
&\tilde{A}_{i_k\,1}^{[m+1]} = \left( \begin{matrix} 0 & 0 \\ \tilde{u}_{m+1}^{i_k} & 0 \end{matrix} \right)\,;~~ \tilde{A}_{i_k\,0}^{[m+1]} = \mathbb{1}\,
\\&
\tilde{u}_{m+1}^{i_k} : \prod_{k=1}^n \tilde{u}_{m+1}^{i_k} = \sum_{\{\mathbf{\Omega}\}} \prod_{\substack{k\,:\mathbf{s}^{(m)}_{k}=1 \\ k':\mathbf{s}^{(m+1)}_{k'}=1 }} 
\!\!\!\! u_{m}^{i_k} u_{m+1}^{i_{k'}}\,. 
\label{eq:decomp}
\end{align}
\end{lemma}
\textit{Proof.}
The right hand side of \eq\eqref{eq:l2} becomes a sum of tensor products in all of which the non-identity matrices occupy the same positions. Therefore, given the structure of the $A$ matrices, these tensor products will return matrices with nonzero elements of the form in \eq\ref{eq:prod_res2}, which depend on the specific vector $\mathbf{s}$, in the exact same positions. Consequently, the result of the sum will also be a matrix with the same structure, where the nonzero elements are the sum of the nonzero elements of the initial matrices.
This sum can be equivalently expressed as a product of scalar factors, with one factor for each entry equal to 
$1$ in $\mathbf{r}$. Furthermore, using Property \ref{prop:tp} again, we can rewrite the sum of tensor products as a single tensor product, where the factors are matrices with the same structure as the original ones, but whose nonzero elements are determined by the decomposition in \eq\eqref{eq:tens_prod_res} of the sum expressed in product form.
\qed

In terms of graphs, this corresponds to the merging of two lines, as it is defined in the Property \ref{prop:merg}.
In the computation of the permanent in the main text, we have, by construction, $\sum_{i=0}^n \mathbf{s}^{(m)}_i = m$ and $\sum_{i=0}^n \mathbf{s}^{(m+1)}_i = 1$, which, by definition, implies $\sum_{i=0}^n \mathbf{r}_i = m+1$.
The Lemma \ref{lem:2} guarantees that the contraction of two $\mathcal{A}$ matrices yields a sum of tensor products, where each term corresponds to a distinct binary vector $\mathbf{r}$. Given that the number of binary vectors of length $n$ containing $m+1$ ones is $\binom{n}{m+1}$, this is the number of tensor products in the resulting sum.
Notably, each step of the decomposition \eq\eqref{eq:decomp} involves a single division and the increment of a sum by one element, thus it can be computed with number of operations $\mathcal{O}(n)$.
\\
\\
\section{Generation of Sierpi\'nski triangles out of $\mathcal{A}$ tensors}
\label{app:sy}
The single photon bosonic operator-basis tensor network can be turned into a standard Fock-basis tensor network through \eq\eqref{eq:matrix_notation},
by expressing the $A$-matrices in terms of the canonical basis of $\mathcal{M}_{2\times 2}$, then by replacing each component with the corresponding $d\times d$ matrix, where $d$ is the local dimension cutoff of bosonic operators.
In the multi-photon case we can perform the same operation starting from the $\mathcal{A}$ tensors. In both cases we can isolate the tensor corresponding to a specific photon number $n_j$ by setting to zero all components featuring a power of $\Aod{j}$ different from $n_j$.
While this approach can construct the target tensors, such procedure is highly inefficient as it relies on the generation of all the components followed by postselection, nullifying the advantage of the method compared to ordinary MPS. 
In this section we show in details how we can overcome this problem by constructing the structure of an $\mathcal{A}$-tensor for a generic number of photons as shown in \fig\ref{fig:fig3}.
Given the enumerated collection $S$ of all $n$ coefficients $u_m^i$ involved in constructing the target $\mathcal{A}$-tensor, we generate the corresponding power set. To order the power set, we rely on its isomorphism with the binary representations of integers in the range $\{0,\ldots,2^n-1\}$. In the case represented in \fig\ref{fig:fig3}, $S=(u_m^i\,u_m^j\,u_m^k)$ and its ordered power set is $\pi(S)=\{() ,(u_m^k), (u_m^j), (u_m^j u_m^k), (u_m^i),(u_m^i u_m^k), (u_m^i u_m^j),(u_m^i u_m^j u_m^k) \}$.
The set difference or relative complement between the sets $A$ and $B$, is defined as $A\backslash B=\{x \in A : x\notin B\}$. Note that the empty set $()$ is included in $\pi(S)$ and $()\backslash A = \varnothing ~~ \forall A \neq ()$ and $A\backslash () = A$.
As with many logical functions \cite{denis1997fractal}, the truth table of the relative complement over the elements of $\pi(S)$ generates a Sierpi\'nski gasket pattern depicted in Fig. \ref{fig:fig3}, which perfectly reproduces the structure of the tensors in \eq\eqref{eq:example_syer}.
Furthermore, the regularity of this structure facilitates the identification and construction of both the coefficients and their positions within the tensor for all blocks that share the same power of creation operators, without the need of generating the whole matrix, simplifying the conversion between the bosonic operator-basis MPS and the standard MPS.

\section{Example: $4$-photon interference}

Here we explicitly employ our formalism to compute the probability amplitude of observing a specific output configuration in a four-photon interference experiment in the defectless scenario. 
This computation of the permanent uses the same operator-basis MPS contractions and tensor network representation described in the main paper, illustrating a concrete application of the formalism.

The input state is $\hat{a}^\dag_1 \hat{a}^\dag_2 \hat{a}^\dag_3 \hat{a}^\dag_4 | {\rm vac}\rangle$ evolves through the unitary optical circuit into a superposition of Fock states and 
we compute the probability of detecting the state
$\hat{a}^\dag_1 \hat{a}^\dag_3 \hat{a}^\dag_7 \hat{a}^\dag_8 | {\rm vac}\rangle$ through photon-counting measurement on the output state. According to \eq (4)
the target amplitude corresponds to the product
\begin{align}
\alpha_{\bar{n}} = \mathcal{A}_{\bar{i}}^{[1]\, 1} \mathcal{A}_{\bar{i}}^{[3]\, 1} \mathcal{A}_{\bar{i}}^{[7]\, 1} \mathcal{A}_{\bar{i}}^{[8]\, 1}\,,
\label{eq:mpo_example}
\end{align}
where $\bar{n}=(1,3,7,8)$ and $\bar{i}=(1,2,3,4)$ and 

\begin{align}
\mathcal{A}_{\bar{i}}^{[j]\, 1}  
\!\!
= 
\!\!\!\!\!\!\!\!\!\!\!
\sum_{\sigma^{(j)}_1 +\ldots + \sigma^{(j)}_4 = 1} \!\!\!\!\!\!\!\!\!\!\!
A_{1\,\sigma^{(j)}_1}^{[j]} \otimes A_{2\, \sigma^{(j)}_2}^{[j]} \otimes A_{3\,\sigma^{(j)}_3}^{[j]} \otimes A_{4\,\sigma^{(j)}_4}^{[j]} \,.
\label{eq:expansion_example}
\end{align}
We start by computing the matrix product 
\begin{widetext}
\begin{align}
\mathcal{A}_{\bar{i}}^{[1]\, 1} \mathcal{A}_{\bar{i}}^{[3]\, 1} =& 
(
A_{1\,1}^{[1]} \otimes A_{2\,0}^{[1]} \otimes A_{3\,0}^{[1]} \otimes A_{4\,0}^{[1]}
+
A_{1\,0}^{[1]} \otimes A_{2\,1}^{[1]} \otimes A_{3\,0}^{[1]} \otimes A_{4\,0}^{[1]}
+
A_{1\,0}^{[1]} \otimes A_{2\,0}^{[1]} \otimes A_{3\,1}^{[1]} \otimes A_{4\,0}^{[1]}
+
A_{1\,0}^{[1]} \otimes A_{2\,0}^{[1]} \otimes A_{3\,0}^{[1]} \otimes A_{4\,1}^{[1]}
) 
\notag\\\times& 
(
A_{1\,1}^{[3]} \otimes A_{2\,0}^{[3]} \otimes A_{3\,0}^{[3]} \otimes A_{4\,0}^{[3]}
+
A_{1\,0}^{[3]} \otimes A_{2\,1}^{[3]} \otimes A_{3\,0}^{[3]} \otimes A_{4\,0}^{[3]}
+
A_{1\,0}^{[3]} \otimes A_{2\,0}^{[3]} \otimes A_{3\,1}^{[3]} \otimes A_{4\,0}^{[3]} +
A_{1\,0}^{[3]} \otimes A_{2\,0}^{[3]} \otimes A_{3\,0}^{[3]} \otimes A_{4\,1}^{[3]}
)
\,,
\label{eq:step1}
\end{align}
\end{widetext}
which, due to the nilpotence of the $A$-matrices, yields
\begin{widetext}
\begin{align}
\mathcal{A}_{\bar{i}}^{[1]\, 1} \mathcal{A}_{\bar{i}}^{[3]\, 1} =& 
(
A_{1\,1}^{[1]} \otimes A_{2\,1}^{[3]} \otimes A_{3\,0}^{[1]} \otimes A_{4\,0}^{[1]}
+
A_{1\,1}^{[1]} \otimes A_{2\,0}^{[1]} \otimes A_{3\,1}^{[3]} \otimes A_{4\,0}^{[1]}
+
A_{1\,1}^{[1]} \otimes A_{2\,0}^{[1]} \otimes A_{3\,0}^{[1]} \otimes A_{4\,1}^{[3]}
+
A_{1\,1}^{[3]} \otimes A_{2\,1}^{[1]} \otimes A_{3\,0}^{[1]} \otimes A_{4\,0}^{[1]}
\notag\\+&
A_{1\,0}^{[1]} \otimes A_{2\,1}^{[1]} \otimes A_{3\,1}^{[3]} \otimes A_{4\,0}^{[1]}
+
A_{1\,0}^{[1]} \otimes A_{2\,1}^{[1]} \otimes A_{3\,0}^{[1]} \otimes A_{4\,1}^{[3]}
+
A_{1\,1}^{[3]} \otimes A_{2\,0}^{[1]} \otimes A_{3\,1}^{[1]} \otimes A_{4\,0}^{[1]}
+
A_{1\,0}^{[1]} \otimes A_{2\,1}^{[3]} \otimes A_{3\,1}^{[1]} \otimes A_{4\,0}^{[1]}
\notag\\+&
A_{1\,0}^{[1]} \otimes A_{2\,0}^{[1]} \otimes A_{3\,1}^{[1]} \otimes A_{4\,1}^{[3]}
+
A_{1\,1}^{[3]} \otimes A_{2\,0}^{[1]} \otimes A_{3\,0}^{[1]} \otimes A_{4\,1}^{[1]}
+
A_{1\,0}^{[1]} \otimes A_{2\,1}^{[3]} \otimes A_{3\,0}^{[1]} \otimes A_{4\,1}^{[1]}
+
A_{1\,0}^{[1]} \otimes A_{2\,0}^{[1]} \otimes A_{3\,1}^{[3]} \otimes A_{4\,1}^{[1]}
)\,,
\label{eq:step3}
\end{align}
\end{widetext}

where, with an abuse of notation, we set $A_{i\,0}^{[k]}\times A_{i\,1}^{[k']} =A_{i\,1}^{[k']} = \left( \begin{matrix} u^{i}_{k'} & 0 \end{matrix} \right)$.
Each term in the sum above is an horizontal vector with a single non-zero element given by the product of the $u^i_j$ coefficients corresponding to each $A^{[j]}_i$ matrix.
Analogously, we can compute the product of the last two tensors
\begin{widetext}
\begin{align}
\mathcal{A}_{\bar{i}}^{[7]\, 1} \mathcal{A}_{\bar{i}}^{[8]\, 1} =& 
(
A_{1\,1}^{[7]} \otimes A_{2\,1}^{[8]} \otimes A_{3\,0}^{[7]} \otimes A_{4\,0}^{[7]}
+
A_{1\,1}^{[7]} \otimes A_{2\,0}^{[7]} \otimes A_{3\,1}^{[8]} \otimes A_{4\,0}^{[7]}
+
A_{1\,1}^{[7]} \otimes A_{2\,0}^{[7]} \otimes A_{3\,0}^{[7]} \otimes A_{4\,1}^{[8]}
\notag\\+&
A_{1\,1}^{[8]} \otimes A_{2\,1}^{[7]} \otimes A_{3\,0}^{[7]} \otimes A_{4\,0}^{[7}
+
A_{1\,0}^{[7]} \otimes A_{2\,1}^{[7]} \otimes A_{3\,1}^{[8]} \otimes A_{4\,0}^{[7]}
+
A_{1\,0}^{[7]} \otimes A_{2\,1}^{[7]} \otimes A_{3\,0}^{[7]} \otimes A_{4\,1}^{[8]}
\notag\\+&
A_{1\,1}^{[8]} \otimes A_{2\,0}^{[7]} \otimes A_{3\,1}^{[7]} \otimes A_{4\,0}^{[7]}
+
A_{1\,0}^{[7]} \otimes A_{2\,1}^{[8]} \otimes A_{3\,1}^{[7]} \otimes A_{4\,0}^{[7]}
+
A_{1\,0}^{[7]} \otimes A_{2\,0}^{[7]} \otimes A_{3\,1}^{[7]} \otimes A_{4\,1}^{[8]}
\notag\\+&
A_{1\,1}^{[8]} \otimes A_{2\,0}^{[7]} \otimes A_{3\,0}^{[7]} \otimes A_{4\,1}^{[7]}
+
A_{1\,0}^{[7]} \otimes A_{2\,1}^{[8]} \otimes A_{3\,0}^{[7]} \otimes A_{4\,1}^{[7]}
+
A_{1\,0}^{[7]} \otimes A_{2\,0}^{[7]} \otimes A_{3\,1}^{[8]} \otimes A_{4\,1}^{[7]}
)\,,
\label{eq:step3}
\end{align}
\end{widetext}
and with the final contraction we obtain
\begin{align}
\alpha_{\bar{n}}&= 
u^{1}_{1}  u^{2}_{3} u^{3}_{7}  u^{4}_{8 }
+
u^{1}_{1}  u^{2}_{3}  u^{3}_{8 }  u^{4}_{7 }
+
u^{1}_{1}  u^{2}_{7}  u^{3}_{3 }  u^{4}_{8 }
+
u^{1}_{1}  u^{2}_{8}  u^{3}_{3 }  u^{4}_{7 }
\notag\\&
+
u^{1}_{1}  u^{2}_{7}  u^{3}_{8 }  u^{4}_{3 }
+
u^{1}_{1}  u^{2}_{8}  u^{3}_{7 }  u^{4}_{3 }
+
u^{1}_{3}  u^{2}_{1}  u^{3}_{7 }  u^{4}_{8 }
+
u^{1}_{3}  u^{2}_{1}  u^{3}_{8 }  u^{4}_{7 }
\notag\\&
+
u^{1}_{7}  u^{2}_{1}  u^{3}_{3 }  u^{4}_{8 }
+
u^{1}_{8}  u^{2}_{1}  u^{3}_{3 }  u^{4}_{7 }
+
u^{1}_{7}  u^{2}_{1}  u^{3}_{8 }  u^{4}_{3 }
+
u^{1}_{8}  u^{2}_{1}  u^{3}_{7 }  u^{4}_{3 }
\notag\\&
+
u^{1}_{3}  u^{2}_{7}  u^{3}_{1 }  u^{4}_{8 }
+
u^{1}_{3}  u^{2}_{8}  u^{3}_{1 }  u^{4}_{7 }
+
u^{1}_{7}  u^{2}_{3}  u^{3}_{1 }  u^{4}_{8 }
+
u^{1}_{8}  u^{2}_{3}  u^{3}_{1 }  u^{4}_{7 }
\notag\\&
+
u^{1}_{7}  u^{2}_{8}  u^{3}_{1 }  u^{4}_{3 }
+
u^{1}_{8}  u^{2}_{7}  u^{3}_{1 }  u^{4}_{3 }
+
u^{1}_{3}  u^{2}_{7}  u^{3}_{8 }  u^{4}_{1 }
+
u^{1}_{3}  u^{2}_{8}  u^{3}_{7 }  u^{4}_{1 }
\notag\\&
+
u^{1}_{7}  u^{2}_{3}  u^{3}_{8 }  u^{4}_{1 }
+
u^{1}_{8}  u^{2}_{3}  u^{3}_{7 }  u^{4}_{1 }
+
u^{1}_{7}  u^{2}_{8}  u^{3}_{3 }  u^{4}_{1 }
+
u^{1}_{8}  u^{2}_{7}  u^{3}_{3 }  u^{4}_{1 }\,,
\label{eq:step4}
\end{align}
which, as shown in \cite{aaronson2011computational}, is the permanent of the matrix
\begin{align}
\mathcal{U}_{\bar{i},\bar{n}} =
\left(
\begin{matrix}
u^1_1 & u^2_1 & u^3_1 & u^4_1\\
u^1_3 & u^2_3 & u^3_3 & u^4_3\\
u^1_7 & u^2_7 & u^3_7 & u^4_7\\
u^1_8 & u^2_8 & u^3_8 & u^4_8\\
\end{matrix}
\right)\,.
\end{align}
The same result can be obtained more easily by using the graphical notation presented in the main text.
The target probability amplitude is given by the contractions in the following graph according to the rules presented in the main text
\begin{align}
\alpha_{\bar{n}} = 
\begin{matrix}
\is & 1 & 2 & 3 & 4 \\
1 &[\, \gs & \ws & \ws & \ws\,]\\
3 &[\, \ws & \gs & \ws & \ws\,]\\
7 &[\, \ws & \ws & \gs & \ws\,]\\
8 &[\, \ws & \ws & \ws & \gs\,]\\
\end{matrix} \,
\label{eq:graph_example}
\end{align}
where we have put $\bar{i}$ and $\bar{n}$ on the top and on the left of the graph respectively.
Merging the first couple of rows generates the following (summed up) configurations, which we report together with the coefficients $\tilde{u}$ corresponding to the $\tilde{A}$-matrices computed according to \eq (A12)
\begin{align} 
&\begin{matrix}
\is & 1 & 2 & 3 & 4 & \is & \is\\
1,3 & \gs & \gs & \ws & \ws & \is &(1+u^1_1 u^2_3/ u^2_1 u^1_3 ~,~ u^2_1 u^1_3)\\
\end{matrix} 
\notag\\&
\begin{matrix}
1,3 & \gs & \ws & \gs & \ws & \is &(1+u^1_1 u^3_3/ u^3_1 u^1_3 ~,~ u^3_1 u^1_3)\\
\end{matrix}
\notag\\&
\begin{matrix}
1,3 & \gs & \ws & \ws & \gs & \is &(1+u^1_1 u^4_3/ u^4_1 u^1_3 ~,~ u^4_1 u^1_3) \\
\end{matrix}
\notag\\&
\begin{matrix}
1,3 & \ws & \gs & \gs & \ws & \is &(1+u^2_1 u^3_3/ u^3_1 u^2_3 ~,~ u^3_1 u^2_3) \\
\end{matrix}
\notag\\&
\begin{matrix}
1,3 & \ws & \gs & \ws & \gs & \is &(1+u^2_1 u^4_3/ u^4_1 u^2_3 ~,~ u^4_1 u^2_3)\\
\end{matrix}
\notag\\&
\begin{matrix}
1,3 & \ws & \ws & \gs & \gs & \is &(1+u^3_1 u^4_3/ u^4_1 u^3_3 ~,~ u^4_1 u^3_3) \\
\end{matrix}
\, .
\label{eq:graph_example2}
\end{align}
Analogously, for the last couple of rows we obtain
\begin{align} 
&\begin{matrix}
\is & 1 & 2 & 3 & 4 & \is & \is\\
7,8 & \ws & \ws & \gs & \gs & \is &(1+u^3_7 u^4_8 / u^4_7 u^3_8 ~,~ u^4_7 u^3_8)\\
\end{matrix} 
\notag\\&
\begin{matrix}
7,8 & \ws & \gs & \ws & \gs & \is &(1+u^2_7 u^4_8/ u^4_7 u^2_8 ~,~ u^4_7 u^2_8)\\
\end{matrix}
\notag\\&
\begin{matrix}
7,8 & \ws & \gs & \gs & \ws & \is &(1+u^2_7 u^3_8/ u^3_7 u^2_8 ~,~ u^3_7 u^2_8) \\
\end{matrix}
\notag\\&
\begin{matrix}
7,8 & \gs & \ws & \ws & \gs & \is &(1+u^1_7 u^4_8/u^4_7 u^1_8 ~,~ u^4_7 u^1_8) \\
\end{matrix}
\notag\\&
\begin{matrix}
7,8 & \gs & \ws & \gs & \ws & \is &(1+u^1_7 u^3_8/ u^3_7 u^1_8 ~,~ u^3_7 u^1_8)\\
\end{matrix}
\notag\\&
\begin{matrix}
7,8 & \gs & \gs & \ws & \ws & \is &(1+u^1_7 u^2_8/ u^2_7 u^1_8 ~,~ u^2_7 u^1_8) \\
\end{matrix}
\, .
\label{eq:graph_example4}
\end{align}
For each row in \eq\eqref{eq:graph_example2} there is exactly one row in \eq\eqref{eq:graph_example4} that can combine with it without resulting in zero. Each non-zero combination contributes a term to the final result, which is the product of the four $\tilde{u}$ values associated with the two rows. Consequently, the final result can be expressed as follows:
\begin{align}
&\alpha_{\bar{n}}= \notag\\&
=(u^1_1 u^2_3 + u^2_1 u^1_3)(u^3_7 u^4_8 + u^4_7 u^3_8)
+
(u^1_1 u^3_3 + u^3_1 u^1_3)(u^2_7 u^4_8 + u^4_7 u^2_8)
\notag\\&+
(u^1_1 u^4_3 + u^4_1 u^1_3)(u^2_7 u^3_8 + u^3_7 u^2_8)
+
(u^2_1 u^3_3 + u^3_1 u^2_3)(u^1_7 u^4_8 + u^4_7 u^1_8) 
\notag\\&+
(u^2_1 u^4_3 + u^4_1 u^2_3)(u^1_7 u^3_8 + u^3_7 u^1_8)
+
(u^3_1 u^4_3 + u^4_1 u^3_3)(u^1_7 u^2_8 + u^2_7 u^1_8) \,,
\end{align}
which corresponds to the permanent in \eq \eqref{eq:step4}. {In both the purely algebraic and graphical computation, we observe that the number of operations is $O(n^2 2^n)$}
\\
\section{N{\"a}ive scaling for Permanent}
\label{app:naive}
For the sake of argument we consider the case $n_i=1$ for $i\le n$ and $0$ otherwise.
Using the mixed-product property of the Kronecker product over all the matrix products in \eq(5)
yields
\begin{align}
&\mathcal{A}_{\bar{i}}^{[1]\, n_1} \mathcal{A}_{\bar{i}}^{[2]\, n_2}\ldots \mathcal{A}_{\bar{i}}^{[M]\, n_M}
\notag
\\&= \!\!\!\!\!\!\!\! \sum_{\sum_{k=1}^n \sigma^{(1)}_k = n_1} \!\!\ldots \!\! \sum_{\sum_{k=1}^n \sigma^{(M)}_k = n_M} 
\!\!\!\!\!\!\!\!
(A_{i_1\,\sigma^{(1)}_1}^{[1]}  A_{i_1 \, \sigma^{(2)}_1}^{[2]}  \ldots A_{i_1\, \sigma^{(M)}_1}^{[M]} )
\otimes\ldots \notag\\&
~~~~~~~\ldots\otimes
(A_{i_n\,\sigma^{(1)}_n}^{[1]} A_{i_n\, \sigma^{(2)}_n}^{[2]} \ldots A_{i_n\,\sigma^{(M)}_n}^{[M]} )\, .
\label{eq:ulmian}
\end{align}
Due to the nilpotence of the matrices $A^{[m]}_{i\,1}$, in each product of matrices in the equation above there can be at most one non-zero ($1$) $\sigma$ index. Furthermore, for each valid configuration, for a fixed $k$, only one $\sigma^{(k)}_j$ is allowed to be $1$.
Therefore counting the non-zero elements of the sum above corresponds to counting all the possible permutations of the lines of an $n\times n$ identity matrix, that, together with the cost of all the matrix multiplications ($n$) gives the computational cost $n \cdot n!$.

\end{document}